# Investigation of Vortex Structures in Gas-Discharge Nonneutral Electron Plasma: IV. Pulse Ejection of Electrons at the mutual interaction of Vortex Structures

## N. A. Kervalishvili

Iv. Javakhishvili Tbilisi State University, E. Andronikashvili Institute of Physics,
6 Tamarashvili Str., Tbilisi 0186, Georgia.   <n_kerv@yahoo.com>

**Abstract.** The results of experimental investigations of the ejection of electrons from gas-discharge nonneutral electron plasma at interaction of vortex structures have been given. The periodical approach of vortex structures causes the ejection of electrons both from the vortex structures themselves and from the adjacent regions of electron sheath to the end cathodes of discharge device. The ejection takes place in the form of short and long pulses following each other. The nature of these pulses and the dynamics of interaction of vortex structures at their approach were studied.

## I. Introduction

In gas-discharge nonneutral electron plasma practically there always present the compact vortex structures with high electron density stretched along the magnetic field [1-4]. To a great extent they determine the properties of such plasma and initiate the different surprising processes taking place in it. One of such processes is the ejection of electrons from the sheath of gas-discharge nonneutral electron plasma to the end cathodes of discharge device. The average value of current of these electrons is quite significant and reaches 50% of the value of discharge current [5, 6]. At the investigation of vortex structures in gas-discharge nonneutral electron plasma it was found that the formation, radial oscillations and approach of vortex structures are accompanied by pulse ejection of electrons from the electron plasma to the end cathodes [1-4].

In paper [7] an electron ejection from the vortex structure during its formation and its radial oscillations at low pressures of neutral gas ($p < 10^{-5} Torr$), when in discharge electron sheath there is only one quasi-stable vortex structure is described in detail. At pressures of the order or more than $5 \times 10^{-5} Torr$ in the magnetron and in the inverted magnetron there simultaneously exist several vortex structures and higher is the pressure the more is their number. The vortex structures move on different orbits with different angular velocities and, therefore, periodically approach each other [1]. The approach of vortex structures is accompanied by pulse ejection of electrons along the magnetic field and such process of electron ejection becomes dominant in gas-discharge nonneutral electron plasma at the pressures of neutral gas of the order or more than $5 \times 10^{-5} Torr$. The present article deals with the study of such process. In paragraph II the process of pulse ejection of electrons at the approach of two vortex structures is investigated. In paragraph III the dynamics of vortex structures when they pass each other is considered and the mechanism of short and long pulses of electron ejection is studied. In paragraph IV the pulse ejection of electrons at other forms of interaction of vortex structures is studied. In paragraph V the variety of forms of pulse electron ejection maintaining the constancy of the density of gas-discharge nonneutral electron plasma in the wide range of pressures of neutral gas is discussed.

## II. Ejection of electrons at the approach of vortex structures

Before starting the analysis of experimental data let us discuss some details of the used experimental methods. These methods were described in detail in [8]. Here, we will consider shortly the possibilities of these methods at the investigation of the motion of two vortex structures. The anode and cathode wall probes, as well as the radial slit in one of the end cathodes are located on one and the same azimuth. This allows to follow the motion of vortex structures past the probes and the motion of the region of electron ejection past the slit. When the vortex structures are far from each other (azimuthally), one can determine their charges and trajectories of their motion, as well as the width of the region of electron ejections from each vortex structure allowing to estimate the transverse dimensions of vortex structures [8]. When the vortex structures approach each other, their orbital trajectories are deformed due to the interactions of vortex structures and therefore, we can observe only the qualitative pattern of the happening. However, even this qualitative pattern appears to be quite sufficient for studying the process of interaction of vortex structures. It should be noted that this method is not fully continuous – it allows to observe the sequential pictures of moving the vortex structure by the wall probes and the slit with the frequency equal to the frequency of the rotation of vortex structure about the axis of discharge device. If the events are developed more slowly than the period of rotation of the vortex structure about of the axis of discharge devise, the process of observing the behavior of vortex structures can be considered to be practically continuous. However, the approach of vortex structures and the event accompanying it are the short-time events. They can take place beyond the region of location of probes and of slit and, in such case, will not be registered by the probes and the slit. Here we can use one more method applied in complete with wall probes and slit. This method is the measurement of the full current of electron ejection to the other end cathode made in the form of screened disc. This method allows to measure continuously the whole process of electron ejection and to determine its duration. Fig.1 gives, as an example, the oscillograms of the current of electron ejection through the slit (upper) and to the disc (lower).

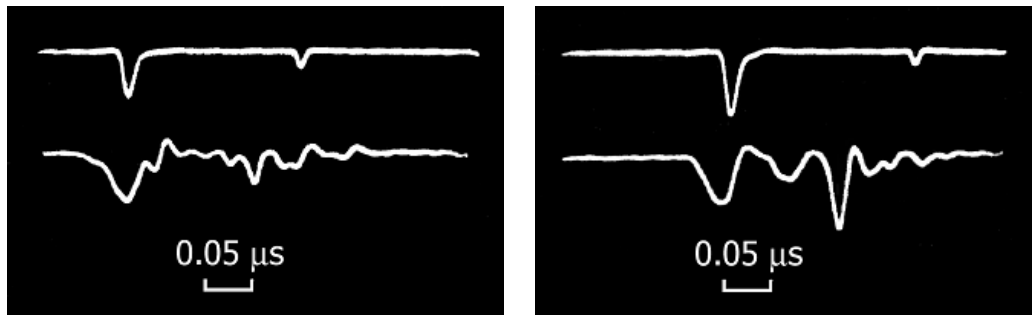

Fig.1. Electron ejection to the disc and through the slit
$r_a = 3.2 cm$; $r_c = 1.0 cm$; $L = 7 cm$; $B = 1.7 kG$; $V = 1.5 kV$; $p = 1 \times 10^{-4} Torr$

In case of short pulse of electron ejection, not only the duration of electron ejection can be determined, but also the azimuth of ejection if comparing the oscillograms of signals from the disc with the oscillograms from the wall probes. In general case, the approach of vortex structures can take place at any azimuth of discharge device. However, for the convenience of interpretation, we chose such oscillograms on which, at least one approach of vortex structures took place at the moment of their passing by wall probes and slit. Moreover, the oscillograms are given in arbitrary and convenient scale.

Fig.2 shows the oscillograms of the current of electron ejection to the end cathodes at approaching two vortex structures. The upper oscillogram shows the oscillations of electric field on the anode wall probe and the lower one – the current of electron ejection to the disc (left) and through the slit (right).

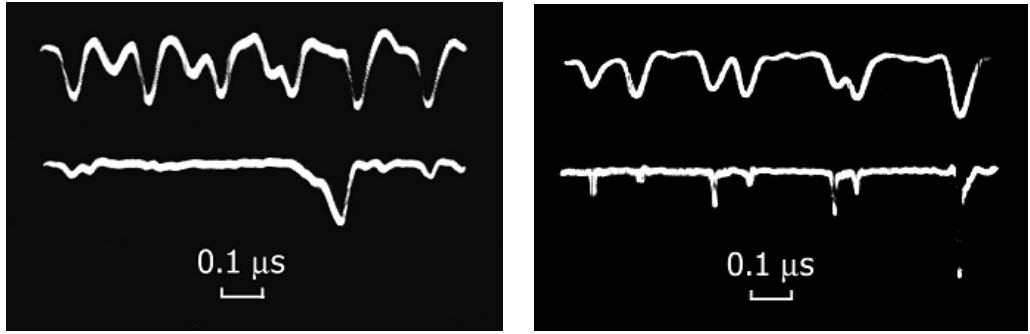

Fig.2 Ejection of electrons at approaching two vortex structures
Left: $r_a = 3.2 cm$; $r_c = 1.0 cm$; $L = 7 cm$; $B = 1.7 kG$; $V = 1.5 kV$; $p = 1 \times 10^{-4} Torr$
Right: $r_a = 3.2 cm$; $r_c = 0$; $L = 7 cm$; $B = 1.9 kG$; $V = 1.0 kV$; $p = 2 \times 10^{-5} Torr$

The approach of vortex structures takes place periodically and each approach is accompanied by pulse electron ejection along the magnetic field [1]. Fig.3 gives the oscillograms of oscillations of electric field on the anode wall probe (upper) and of the electron current on the end cathode (lower) in the case of periodical approach of two vortex structures. As it is seen from the figure, the periodical increase of the amplitude of electric field oscillation takes place on the anode wall probe accompanied by short pulses of electron ejection to the end cathode. In the interval between the short pulses the long pulses of electron ejection are observed, the duration of which is much more than the period of rotation of vortex structure about the axis of discharge device.

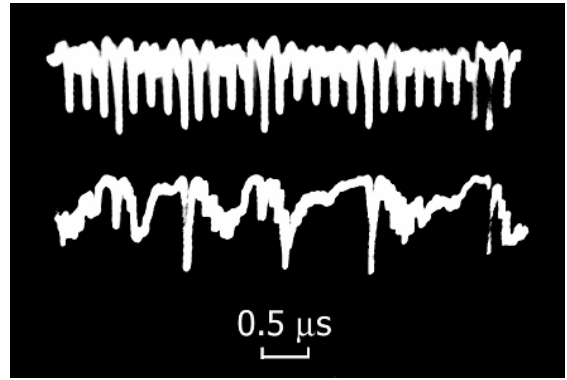

Fig.3 Periodical interaction of vortex structures
$r_a = 3.2 cm$; $r_c = 1.0 cm$; $L = 7 cm$; $B = 1.5 kG$; $V = 2.0 kV$; $p = 5 \times 10^{-5} Torr$

A more detailed illustration of these processes is presented in Figs 4 and 5. Fig.4 shows the oscillograms in the case of interaction of two vortex structures. Fig.5 shows the interaction of vortex structures, when their number is more than two. In both cases, the interaction of vortex structures is accompanied by similar physical processes, however, the interpretation of oscillograms in the case of two vortex structure is easier and more demonstrable. Therefore, in the future we will use the oscillograms for two vortex structures. As it is seen from Fig.4, the vortex structure with higher amplitude of electric field oscillations on the anode wall probe overtakes the vortex structure with lower amplitude of oscillations. The approach of two vortex structures is accompanied by the increase of the amplitude of electric field oscillations on the anode wall probe and by a short pulse of electron ejection to the end cathode.

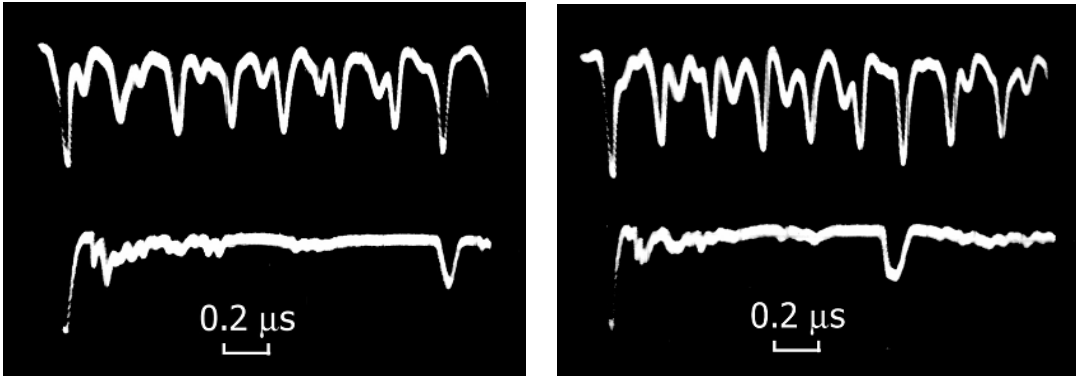

Fig.4 Interaction of two vortex structures
$r_a = 3.2 cm$; $r_c = 1.0 cm$; $L = 7 cm$; $B = 1.5 kG$; $V = 1.5 kV$; $p = 1 \times 10^{-4} Torr$

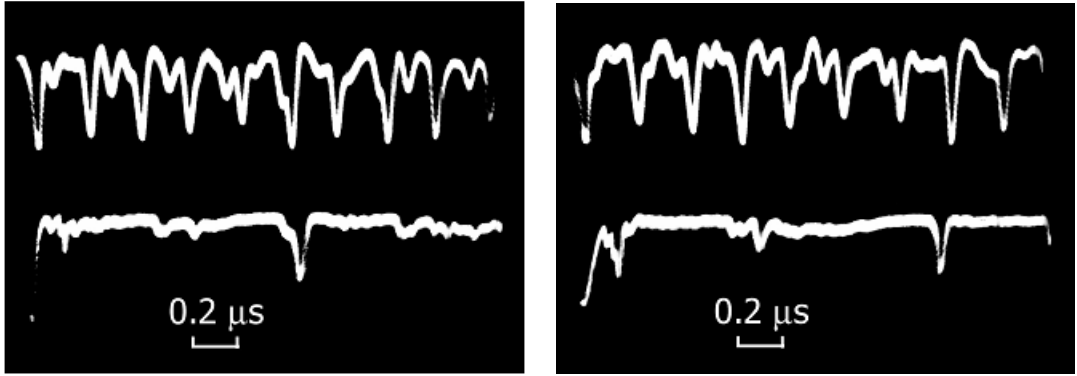

Fig.5 Interaction of several vortex structures
$r_a = 3.2 cm$; $r_c = 1.0 cm$; $L = 7 cm$; $B = 1.5 kG$; $V = 1.5 kV$; $p = 1 \times 10^{-4} Torr$

For more detail information on the charges, orbits and dynamics of vortex structure the additional oscillograms of the oscillations of electric field on the cathode wall probe are required. Fig.6 shows the oscillograms of oscillations of electric field on the anode (upper) and cathode (average) wall probes and of electron current on the end cathode (lower). From the ratio of the amplitudes of electric field oscillations on the anode and cathode wall probes for each of vortex structures it can be concluded that the vortex structure the higher oscillation amplitude corresponds to in Fig.6 has a greater charge, is located near to the anode and moves with high angular velocity. As it is seen from the oscillograms, the approach of vortex structures is accompanied by the increase of the amplitude of electric field oscillations on the anode and cathode wall probes. Synchronous of the pulses of electron ejection with the moment of the approach of vortex structure gives evidence of the fact that the ejection of electrons takes place from the region of electron sheath where the vortex structures approach each other. The short pulse of ejection is followed by a longer pulse continued during several rotations of vortex structures about the axis of discharge device.

The simplest explanation of the pulse electron ejection consists in that at approaching of two vortex structures on the place of their approach the charge of electron sheath is increased locally and, correspondingly, the potential barrier along the magnetic field decreases [1]. The estimation of the decrease of the potential barrier at the approach of two vortex structures presented in [1] gives 0.3-0.7 value of the value of discharge voltage being quite sufficient for ejection of a part of electrons to the end cathodes from the vortex structures and from the adjacent region of electron sheath. However, such mechanism does not explain the origination of long electron pulse.

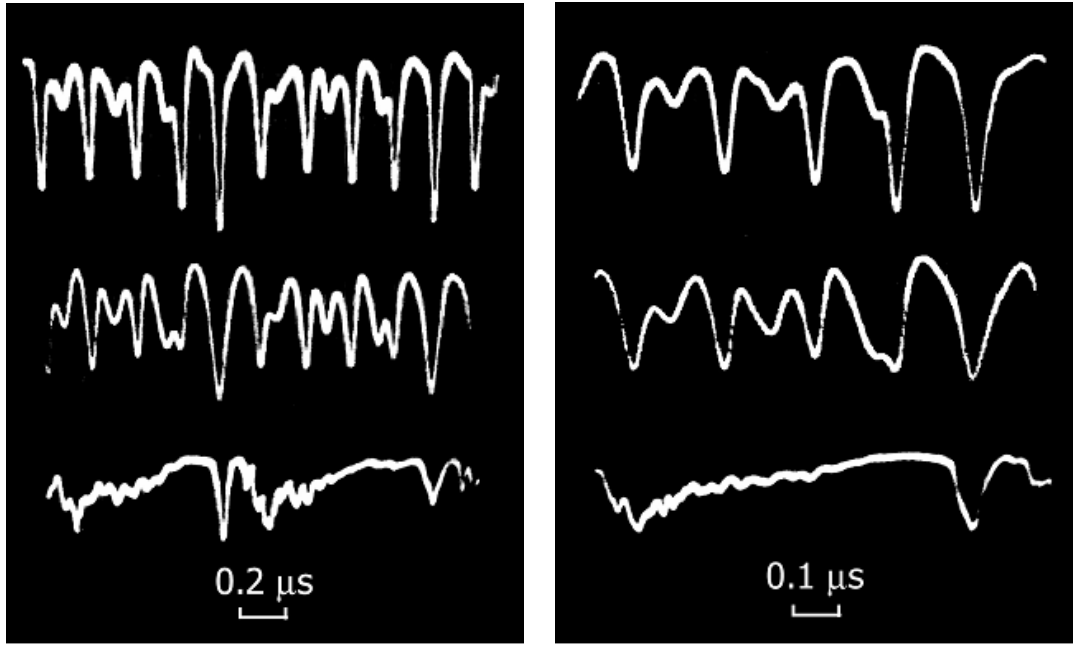

Fig.6 Ejection of electrons at approaching two vortex structures
$r_a = 3.2 cm$; $r_c = 1.0 cm$; $L = 7 cm$; $B = 1.5 kG$; $V = 2.0 kV$; $p = 5 \times 10^{-5} Torr$

**III. The mechanism of electron ejection at the interaction of vortex structures**

Let us consider the approach of two vortex structures in more detail. When the vortex structures are located far from each other each of them moves in a circular orbit about the axis of discharge device. However, each vortex structure has its own excess charge and, hence, has its own electric field. At the approach of vortex structures, each of them makes an additional drift in the electric field of the other vortex structure. The vortex structures bypass each other and move away from each other to the radial direction. The main vortex structure existing near to the anode surface is displaced towards the anode, and second vortex structure removed farther from the anode surface, on the contrary, is displaced off the anode. Thus, the both vortex structures make periodically the forced radial oscillations in contrast to self-induced oscillations of one vortex structure during the orbital instability [1, 7]. The second vortex structure moves away from the anode surface for a short time (the time of approach of vortex structures) and therefore, a short pulse of electron ejection is probably from this vortex structure. The vortex structure located nearer to the anode approaches the anode for a short time, while for a longer time equal to the time interval between the approaches of vortex structures it moves away from the anode surface (returns to its stationary orbit). Probably, the long pulse of electron ejection corresponds to this process. On the oscillograms given in Fig.6 it is seen that the higher amplitude of oscillations of electric field on the anode wall probe corresponds to the vortex structure located nearer to the anode surface. This follows, first of all, from the ratio of the amplitudes of oscillations of electric field on the anode and cathode wall probes for both vortex structures, and, secondly, from the ratio of angular velocities of the rotation of vortex structures about the axis of discharge device. Correspondingly, the long pulse of electron ejection is probably from this vortex structure. The oscillograms shown in Fig.7 confirm such assumption. The upper two oscillograms are the oscillations of electric field on the anode (upper) and cathode (the second from above) wall probes. The low oscillogram is the electron current to the end cathode (disc). The second from the below oscillogram is the electron current to the sector of disc on the other end cathode. The central angle of sector equaled $120^0$ and its center was located in the same azimuth as the anode and cathode wall probes. The disc and the sector were shielded against the electrostatic pickup.

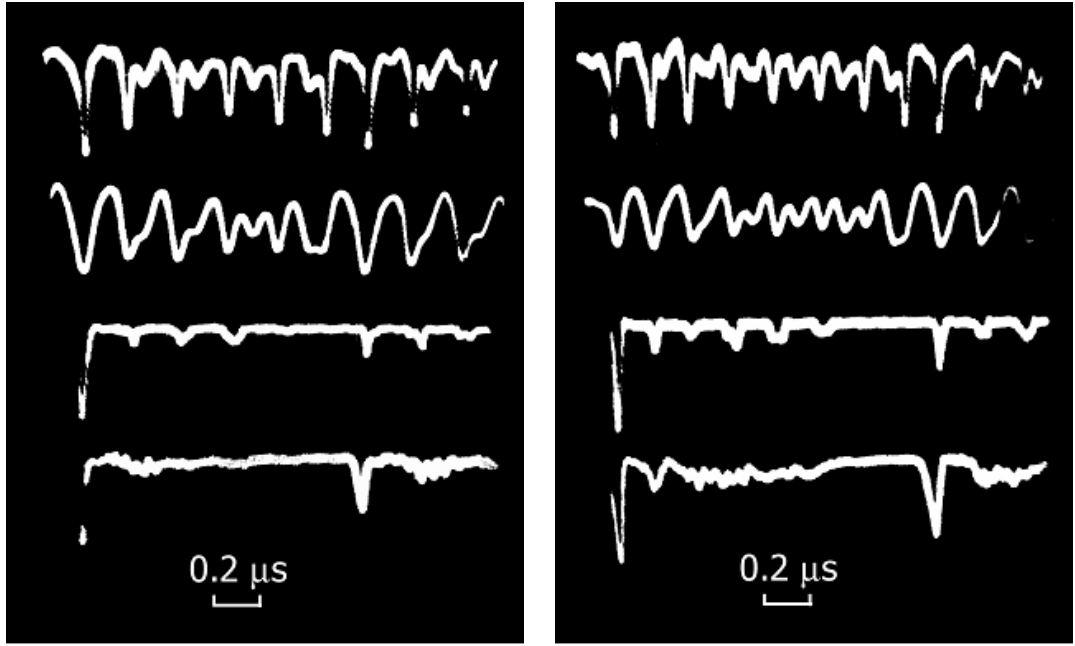

Fig.7 Ejection of electrons to the sector and to the disc at the approach of two vortex structures
Left: $r_a = 3.2 cm$; $r_c = 1.0 cm$; $L = 7 cm$; $B = 1.5 kG$; $V = 1.5 kV$; $p = 1 \times 10^{-4} Torr$
Right: $r_a = 3.2 cm$; $r_c = 1.0 cm$; $L = 7 cm$; $B = 1.5 kG$; $V = 2.0 kV$; $p = 1 \times 10^{-4} Torr$

As it is seen from the figure, in the interval of time between the approaches of vortex structures the pulses on the sector are synchronous with the pulses on the wall probes from the vortex structure located nearer to the anode surface, while the pulses synchronous with the second vortex structure are not at all. Here, like Fig.6 the vortex structure to which the higher amplitude of oscillations on the anode and cathode wall probes corresponds, has a greater charge and is located nearer to the anode. This oscillogram serves as a confirmation of the fact that the long ejection of electrons takes place from the vortex structure located near to the anode and that the region of ejection rotates together with the vortex structure about the axis of discharge device. Here it should be also noted that the duration of long pulse, as it is seen on Figs. 6 and 7 is shorter than the interval of time between the approaches of vortex structures and the right edge of pulse is sloping. This is in agreement with the fact that the electrons acquire the longitudinal velocity at the expense of electron-neutral collisions and, hence, "reserve" of electrons having the longitudinal velocity sufficient to overcome the potential barrier exhausts gradually.

On the oscillograms (Fig.7) the short pulses of electron ejection on the disc and on the sector coincide with the moment of the approach of vortex structures. This indicates that the ejection of electrons takes place, first of all, at the moment of approach of vortex structures and secondly, at the place of their approach. The ejection of electrons, probably, takes place predominantly from the second vortex structure, as at the moment of ejection it displaces from the anode surface. As a confirmation of the ejection of electrons predominantly from the second vortex structure can serve the oscillograms in Fig.8. Here, the upper oscillogram is the oscillations of electric field on the anode wall probe, and the lower one – the current of electron ejection through the radial slit to the end cathode. The vortex structures the oscillogram refers to are located on the orbits close to each other have about the same charges and are approaching slowly. A small ejection of electrons takes place from the both vortex structures. However, at the moment of time directly preceding the final approach of vortex structures the ejection of electrons from the second vortex structure strongly increases. This serves as a confirmation of the fact that at the moment of approach the second vortex structure is displaced from the anode surface.

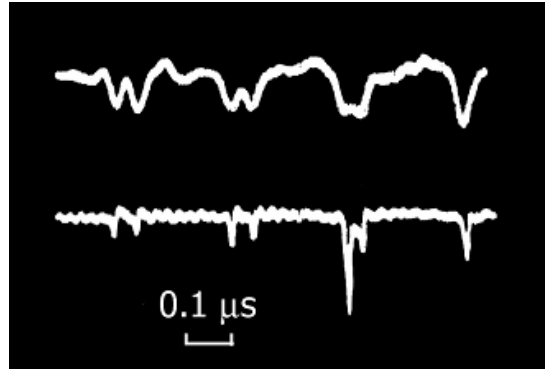

Fig.8 Ejection of electrons through the slit at the approach of two vortex structures
Right: $r_a = 3.2 cm$; $r_c = 0$; $L = 7 cm$; $B = 1.9 kG$; $V = 1.0 kV$; $p = 2 \times 10^{-5} Torr$

**IV Other forms of interaction of vortex structures**

The interaction of vortex structures is not limited only to the process of their periodical approach. There are also the other forms of interaction, e.g. merging. But, even in the case of periodical approach some difference in the dynamics of vortex structures is observed. In the most cases the periods of turn of both vortex structures about the axis of discharge device differ from each other by about 15-20%, i.e. they are located on the orbits close to each other. At the same time, the vortex structure located near to the anode has, as a rule, a greater charge. However, sometimes, the orbits of vortex structures are located rather far from each other and this makes a certain difficulties at their identification by means of wall probes.

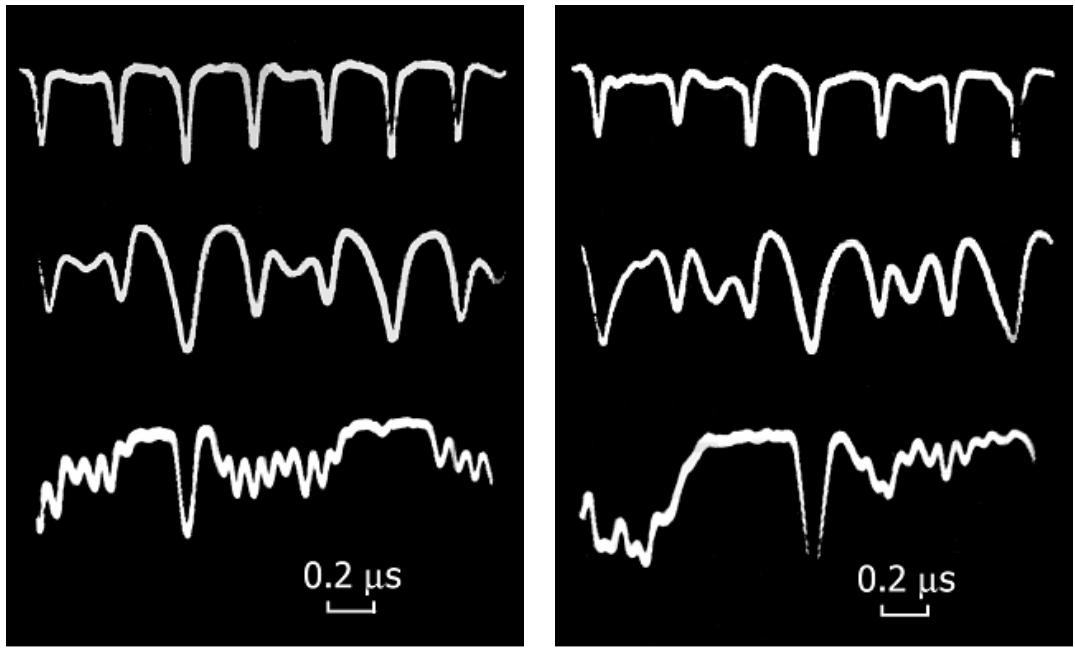

Fig.9. One of the vortex structures is located close to the cathode
$r_a = 3.2 cm$; $r_c = 1.0 cm$; $L = 7 cm$; $B = 1.5 kG$; $V = 1.0 kV$; $p = 5 \times 10^{-5} Torr$

Fig.9 shows the oscillograms of the oscillations of electric field on the anode (upper) and the cathode (average) wall probes, as well as the current of electron ejection to the end cathode (lower). These oscillograms illustrate the periodical approach of two vortex structures the period of the turn of which about axis of discharge device differ from each other by one and a half times. The second vortex structure has a much less charge and is located much closer to the cathode than the main one. It is not practically notable on the oscillogram from the anode wall probe. And only the signal from the cathode wall probe and the pulses of electron ejection to the end cathode allow to judge about the existence of the second vortex structure and about the periodical approach of vortex structures.

Fig.10 gives the oscillograms illustrating the opposite case, when the second vortex structure is located much closer to the anode, has a much less charge and rotates with higher angular velocity, as compared to the main vortex structure. It is practically unnoticeable on the oscillogram from the cathode wall probe (lower), however, it is seen clearly on the oscillogram from the anode wall probe (upper).

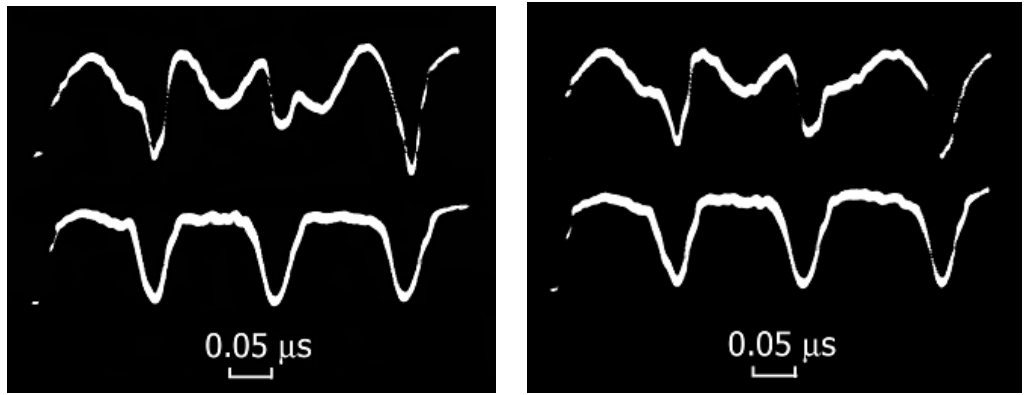

Fig.10. One of the vortex structures is located close to the anode
$r_a = 3.2cm$; $r_c = 1.0cm$; $L = 7cm$; $B = 1.5kG$; $V = 3.0kV$; $p = 5 \times 10^{-5} Torr$

Now let us consider the approach of two vortex structures completed by their merging. The effect is observed most clearly at low pressures of neutral gas during the formation of one of quasi-stable vortex structure [9, 10]. On the left of Fig.11 the oscillograms from the anode (upper) and cathode (lower) wall probes are given showing the process of merging of two vortex structures in the inverted magnetron.

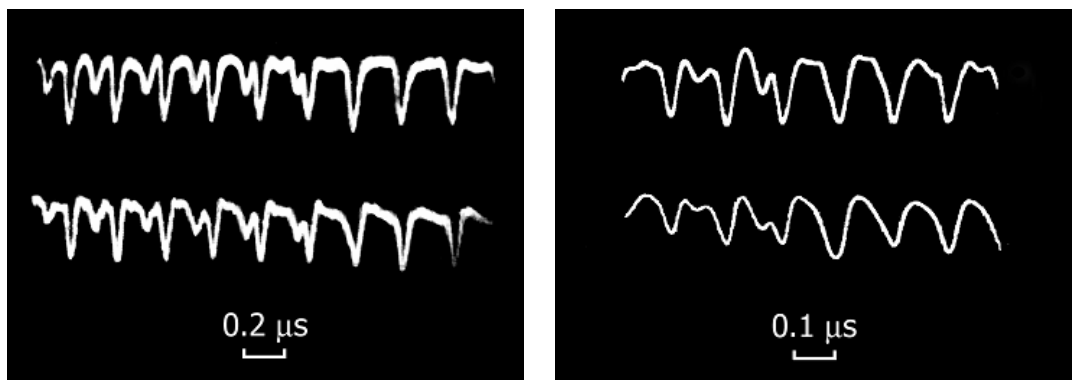

Fig.11 Merger of two vortex structures
Left: $r_a = 2.0cm$; $r_c = 3.2cm$; $L = 7cm$; $B = 1.5kG$; $V = 1.0kV$; $p = 2 \times 10^{-5} Torr$
Right: $r_a = 3.2cm$; $r_c = 1.0cm$; $L = 7cm$; $B = 1.5kG$; $V = 1.5kV$; $p = 2 \times 10^{-4} Torr$

The merging of vortex structures is accompanied, as in the case of usual approach, by a short pulse of electron ejection. After this, the solitary vortex structure continues to exist for a long time – much more than the time of electron-neutral collisions.

At higher pressures of neutral gas when there are simultaneously several vortex structures the cases of merging of vortex structures at their approach are also observed. On the right of Fig.11 the oscillograms from the anode (upper) and cathode (lower) wall probes are given showing the process of merging of two vortex structures in the magnetron geometry. However, after several turns of such "merged" vortex structure about the axis of discharge device, the origination of a new vortex structure takes place the charge of which increases gradually. Such cycle can be periodically repeated. The oscillograms of such process in the magnetron geometry of discharge device are given in Fig.12. Here, the upper oscillogram is the oscillations of electric field on the anode wall probe, and the lower one – the current of electron ejection to the end cathode. As it is seen from the figure, the merging of vortex structures is accompanied by a short pulse of electron ejection.

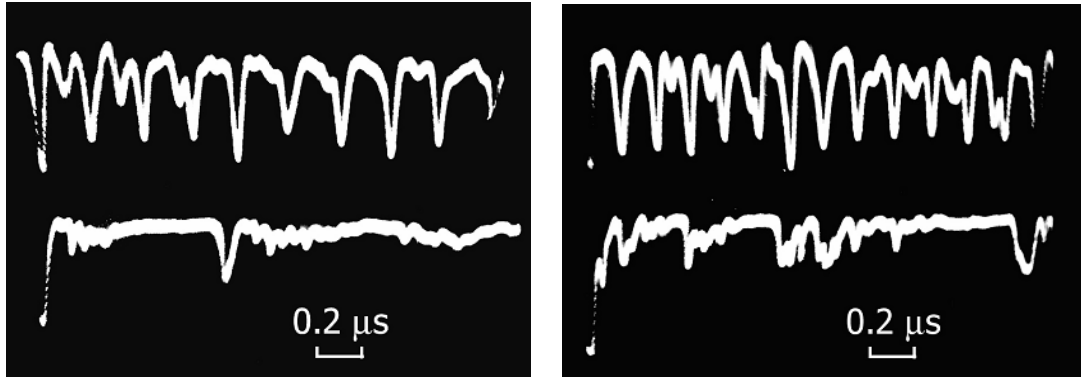

Fig.12 Merging of vortex structures and origination of a new vortex structure
Left: $r_a = 3.2 cm$; $r_c = 1.0 cm$; $L = 7 cm$; $B = 1.5 kG$; $V = 1.5 kV$; $p = 2 \times 10^{-4} Torr$
Right: $r_a = 3.2 cm$; $r_c = 1.0 cm$; $L = 7 cm$; $B = 1.5 kG$; $V = 1.5 kV$; $p = 4 \times 10^{-4} Torr$

The process of merging of two equal vortex structures is experimentally studied in pure electron plasma [11] and it was shown that the merging of vortex structures takes place quite rapidly if the distance between their centers is less than some critical value. Such process of merging can take place as well in our case if the vortex structures are approaching each other quite near at their approach. However, the other explanation of the observed process of merging can also be supposed. As it was mentioned above, in the process of approaching the vortex structures bypass each other and at the same time, move away from each other to the radial direction. The second vortex structure located farther from the anode surface moves even more away from it and at the same time losses a part of electrons. At sufficiently large departure from the anode surface, the loss of electrons can be such great that on the oscillograms from the anode and cathode wall probes the vortex structure can be practically unnoticeable. This can be interpreted as a merge with the main vortex structure. Further on, thanks to the ionization, the charge of the second vortex structure will start to increase. Correspondingly, the signal from this structure to the wall probes will be increased as well and this will be interpreted as an origination of a new vortex structure. Thus, instead of merging and following origination of a new vortex structure, we will have the process of losing of a significant part of the charge and, then, its further accumulation by one and the same vortex structure. Such development of events is evidenced by the fact that the signal on the wall probes from the merged vortex structures differs slightly from the signal from the main vortex structure. On the other hand, the origination of a new vortex structure sometimes is observed as well in the cases when the approaching vortex structures are not merged but simply pass by each other. The example of such interaction of vortex structures is given in Fig.13. Here, two upper oscillograms are the oscillations of electric field on the anode (upper) and cathode (the second from above) wall probes. The lower oscillogram is the current of electron ejection on the end cathode (disc). The second from the below oscillogram is the current of electron ejection on the sector of other end cathode.

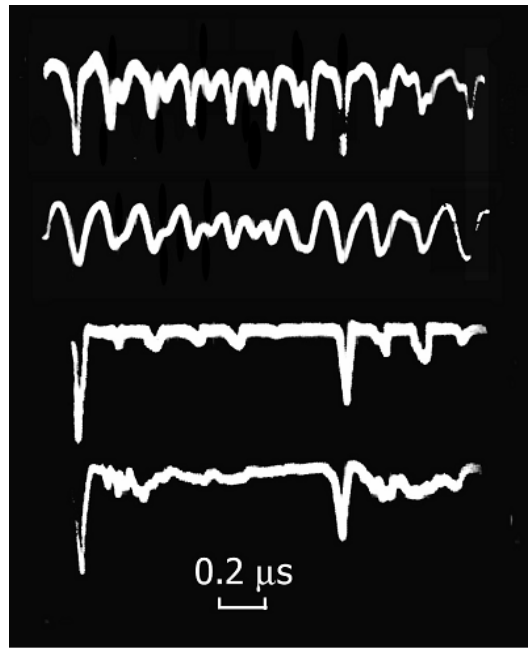

Fig.13 Origination of a new vortex structure at approaching of two vortex structures
$r_a = 3.2 cm$; $r_c = 1.0 cm$; $L = 7 cm$; $B = 1.5 kG$; $V = 2.0 kV$; $p = 2 \times 10^{-4} Torr$

Oscillograms are chosen in such a way that the both approaches of two vortex structures are in the region where the wall probes and the sector are located. The process of origination of a new vortex structure is shown in detail in Fig.14. Here, the upper oscillogram is the oscillations of electric field on the anode wall probe, and the lower one – the current of electron ejection to the sector. From the oscillograms in Fig.13 it follows that the process of approaching of two vortex structures is accompanied by short pulses of electron ejection, and in the interval between the approaches of vortex structures the ejection of electrons takes place from the vortex structure located near to the anode. From Fig.14 it follows that in the process of origination of a new vortex structure the ejection of electrons takes place as well from the main vortex structure located near to the anode.

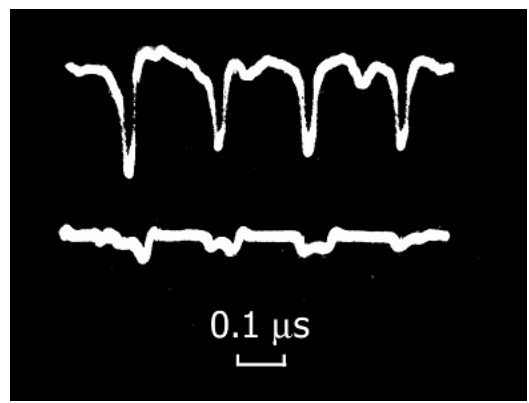

Fig.14 Origination of a new vortex structure
$r_a = 3.2 cm$; $r_c = 1.0 cm$; $L = 7 cm$; $B = 1.5 kG$; $V = 2.0 kV$; $p = 2 \times 10^{-4} Torr$

## V. Conclusion

Possibly, the ejection of electrons from vortex structures and adjacent regions of electron sheath to the end cathodes of discharge device is one of the fundamental properties of gas-discharge nonnetrual electron plasma. In the present work the pulse ejection of electrons taking place as a result of interaction of vortex structures was studied. Earlier, [7] the pulse ejection of electrons taking place at the formation and at radial oscillations of solitary quasi-stable vortex structure was considered. However, despite the difference of processes of electron ejection, they have one and the same mechanism – short-term local decrease of the potential barrier along the magnetic field. The local decrease of the potential barrier takes place either at the expense of the local increase of electron density, or at the expense of displacement of the vortex structure to the region with lower potential barrier. The electrons that acquired the "necessary" longitudinal velocity at the expense of collisions with neutral atoms at high potential barrier and, correspondingly, at high average energy of electrons, leave the electron sheath at the decreased potential barrier.

In order the electrons could gain the necessary longitudinal velocity at the expense of electron-neutral collisions, the frequency of approaching two vortex structures should be less or of the order of the frequency of electron-neutral collisions. Such condition is fulfilled at relatively high pressures of neutral gas. At further increase of the pressure of neutral gas, the intensity of ionization increases and, simultaneously, increases the number of vortex structures leading to more frequent approach of vortex structures. Thus, the increase of the reproduction of electrons at the expense of the increase of the frequency of ionization is compensated by the increase of electron ejection at the expense of the increase of the number of vortex structures. Otherwise, at the decrease of the pressure of neutral gas the frequency of ionization decreases and the frequency of the approach of two vortex structures, practically, is not changed. In this case, the amplitude of electron ejection at each following approach of vortex structures starts to decrease. Finally, the vortex structures merge with each other and only one quasi-stable vortex structure is left. Here, the other mechanisms of electron ejection are involved [7]. Thus, the vortex structures formed in gas-discharge nonneutral electron plasma initiate the various processes of electron ejection and in this way keep the density of electron plasma unchanged in the wide range of neutral gas pressures.